\documentclass[aps,twocolumn,preprintnumbers,superscriptaddress,
              showpacs,nofootinbib]{revtex4}
 \newcommand{\PRE}[1]{}       
\usepackage{amsmath,amsfonts,epsfig,graphicx}
\usepackage{multirow}
\usepackage[latin1]{inputenc}
 \usepackage{amssymb}
 \usepackage{verbatim}
\usepackage{float}
\usepackage{slashed} 

\newcommand{\met}{\ensuremath{\slashed{E}_T}}

\newcommand{\MGMCatNLO}{MadGraph5\_aMC@NLO} 
\newcommand{\madgraph}{{\sc MadGraph}}
\newcommand{\decay}{{\sc DECAY}}

\newcommand{\pythia}{{\sc Pythia}}

\newcommand{\delphes}{{\sc Delphes}}

\begin{document}

\title{Polarization fraction measurement in same-sign $WW$ scattering using deep learning}

\author{Junho Lee}
\affiliation{
Department of Physics and State Key Laboratory of Nuclear Physics and Technology, Peking University, Beijing, 100871, China}

\author{Nicolas Chanon}
\affiliation{
Institut de Physique Nucl\'{e}aire de Lyon, Universit\'{e} de Lyon, Universit\'{e} Claude Bernard Lyon 1, CNRS-IN2P3, Villeurbanne 69622, France }

\author{Andrew Levin}
\affiliation{
Department of Physics and State Key Laboratory of Nuclear Physics and Technology, Peking University, Beijing, 100871, China}

\author{Jing Li}
\affiliation{
Department of Physics and State Key Laboratory of Nuclear Physics and Technology, Peking University, Beijing, 100871, China}

\author{Meng Lu}
\affiliation{
Department of Physics and State Key Laboratory of Nuclear Physics and Technology, Peking University, Beijing, 100871, China}

\author{Qiang Li}
\affiliation{
Department of Physics and State Key Laboratory of Nuclear Physics and Technology, Peking University, Beijing, 100871, China}

\author{Yajun Mao}
\affiliation{
Department of Physics and State Key Laboratory of Nuclear Physics and Technology, Peking University, Beijing, 100871, China}

\preprint{VBSCAN-PUB-09-18}

\begin{abstract}
Studying the longitudinally polarized fraction of $W^\pm W^\pm$ scattering at the LHC is crucial to examine the unitarization mechanism of the vector boson scattering amplitude through Higgs and possible new physics. We apply here for the first time a Deep Neural Network classification to extract the longitudinal fraction. Based on fast simulation implemented with the Delphes framework, significant improvement from a deep neural network is found to be achievable and robust over all dijet mass region. A conservative estimation shows that a high significance of four standard deviations can be reached with the High-Luminosity LHC designed luminosity of 3000 $fb^{-1}$

\end{abstract}


\maketitle

The High-Luminosity LHC (HL-LHC)  will measure for the first time many novel processes predicted by standard model (SM), and study precisely especially those involving pure electroweak interactions such as vector boson scattering (VBS). VBS is sensitive to non-Abelian weak gauge boson interactions, and to the structure of electroweak symmetry breaking. Typical VBS signatures at hadron colliders include, for example, large dijet mass ($m_{\text{jj}}$) and large pseudorapidity separation ($\Delta \eta_{jj}$).

Among various VBS processes, same charge $W^\pm W^\pm$ production is one of the most promising channels for the above mentioned purpose. The VBS $W^\pm W^\pm$ process profits from low background, due to the signature of two same sign charged leptons. Same charge $W^\pm W^\pm$ scattering has been observed by CMS and ATLAS with a significance larger than 5 standard deviations, based on data collected at $\sqrt{s}=13$~TeV, corresponding to an integrated luminosity of approximately $35.9fb^{-1}$~\cite{Sirunyan:2017ret}~\cite{ATLAS:2018ogo}. The dominant backgrounds after the VBS selection arise from WZ production with one lepton misidentified, and non-prompt leptons from hadron decays, which can be further suppressed by requiring $m_{\text{jj}}$ to be above 1 TeV.

The next important goal after the discovery of VBS $W^\pm W^\pm$ is to measure the fraction of longitudinally polarized (LL) events. The LL component contributes only to a level of 5-10\% in $W^\pm W^\pm \rightarrow W^\pm W^\pm$, but it is extremely interesting as a direct probe of the unitarization mechanism~\cite{WWuni} of the vector boson scattering amplitude through Higgs and possible new physics~\cite{Alboteanu:2008my} ~\cite{Chang:2013aya}.

There have been extensive studies on LL fraction measurement, exploiting various kinematic observables. Popular variables include leading lepton transverse momentum ($p_T^{~l_1}$), and the azimuthal angle difference between the two leading jets ($\Delta\phi_{\text{jj}}$). On top of these, ref.~\cite{Doroba12} proposed to use the variable $R_{p_T} = {p_T^{~l_1}\cdot p_T^{~l_2}\over p_T^{~j_1}\cdot p_T^{~j_2}}$. Ref.~\cite{Freitas:2012uk} examined matrix element method to differentiate different beyond SM model scenarios. More recently, ref.~\cite{Searcy:2015apa} applied a regression with Deep Neural Network (DNN) to recover the lepton angular distributions in the W boson rest frame, and shows that the expected accuracy can be improved by about a factor of two compared to the use of $R_{p_T}$.

In the meantime, CMS studied the prospects for a measurement of the LL fraction, based on full simulation samples with the upgraded CMS detector at the 14 TeV HL-LHC~\cite{CMS-PAS-SMP-14-008,CMS-PAS-FTR-18-005}. The expected significance for an integrated luminosity of 3000 $fb^{-1}$ is estimated to be 2.7 standard deviations. The study is based on a fit to $\Delta\phi_{\text{jj}}$ distributions in two $m_{\text{jj}}$ bins.

In this study, we examine the impact of using a DNN on LL fraction measurement. In contrast to what has been done in ref.~\cite{Searcy:2015apa}, we exploit here DNN classification instead of regression, based on the framework of the Keras library~\cite{Keras} with Tensorflow back-end~\cite{Tensorflow}. We perform a fit on the resulting DNN discriminant.

There have been more and more applications of machine learning techniques in high energy physics, with some first examples in Refs~\cite{Baldi:2014kfa, Baldi:2014pta}. Detailed studies are provided in this paper based on either low-level or high-level features. A comparison with boosted-decision trees~\cite{Roe:2004na} implemented in TMVA~\cite{TMVA} are also provided. 

Simulation samples are generated with {\MGMCatNLO}~\cite{Alwall:2014hca} interfaced with \pythia\,6~\cite{Sjostrand:2003wg} for parton showering and hadronization and \delphes~version 3~\cite{deFavereau:2013fsa} for detector simulation with CMS configuration. Similarly as in ref.~\cite{Searcy:2015apa}, we neglect the `pileup' effects due to overlapping interactions in proton proton collision, as they can be mitigated effectively with advanced experimental techniques. The inclusive $W^\pm W^\pm$ VBS samples are decomposed into LL, TT (transversely polarized $W^\pm W^\pm$) and TL (transversly and longitudinally polarized $W^\pm W^\pm$) components, with the help of \decay\ package provided by \madgraph.  We require exactly 2 same-sign charged leptons with $p_T>20$~GeV and $|\eta|<2.4$, and select the two leading jets with $p_T>50$~GeV and $|\eta|<4.7$ as VBS jet candidates. We further require $|\Delta\eta_{\text{jj}}|>2.5$, a b jet veto, and performed our studies in several benchmark selections : $m_{\text{jj}} > $ 850, 1200, 1500, 1800, and 2000 GeV. The backgrounds from WZ and non-prompt leptons can indeed be suppressed effectively with higher $m_{\text{jj}}$.

As inputs to the DNN, low-level features include the $p_T$, $\eta$, $\phi$ of the two leptons, $p_T$, $\eta$, $\phi$ and mass of the two jets, and x- and y- components of missing transverse energy (\met). We further include high-level features with zeppenfeld variable~\cite{Rainwater:1996ud} of the two leptons, $\Delta\phi_{\text{jj}}$, $\Delta\eta_{\text{jj}}$ and $\Delta{\text{R}}_{\text{ll,jj}}$. Four million events have been produced for training, validation and testing. Overtraining has been carefully checked by monitoring loss value dependency on DNN training epoch, in both the training and validation dataset. Early stopping has been applied if there is no improvement in loss value comparing with any latest 20 epoch's loss value. Overtraining can also be precisely checked by comparing output distribution of training and test dataset.   

\begin{figure}[htbp]
  \begin{center}
  \includegraphics[width=0.4\textwidth]{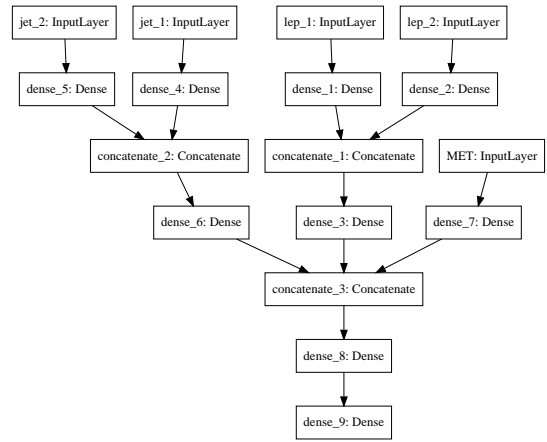}
    \caption{Simplified structure of the particle-based DNN model. Inputs are features of each particle, gradually merging to the output layers.}
    \label{fig:dnn_model}
  \end{center}
\end{figure}

\begin{figure}[htbp]
  \begin{center}
   \includegraphics[width=0.4\textwidth]{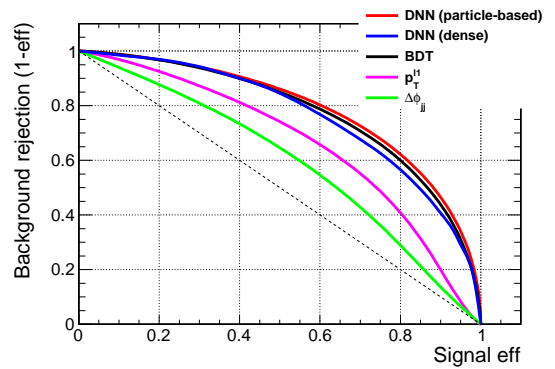}
    \caption{ROC curves with a selection of $m_{\text{jj}}>1500$~GeV applied to VBS $W^\pm W^\pm$. The X-axis showing signal efficiency of LL component, and Y-axis the rejection rate of TT+TL components}
    \label{fig:roc}
  \end{center}
\end{figure}

Two differently structured DNN models, a `dense' and a `particle-based' model, have been trained and tested. We selected a 10-layers dense neural network with 150 hidden units on each layer with the `relu' activation function, the `sigmoid' function applied on final nodes, taking `adam' optimizer with a learning rate of 0.001, and 0.01 as regularization term for L2 regularization. Moreover, a batch size of 50 events, a 50\% drop-out rate on hidden unit, and batch normalization are applied to avoid overtraining.
As an alternative, to efficiently model highly correlated variables of each particle, we also tried a particle-based model which involves separate grouping of nodes for the features of each particle and a gradual merging of all nodes into bigger layers. Fig.~\ref{fig:dnn_model} shows a simplified version of the particle-based model. The model actually used contains 2 hidden layers with 20 nodes for each particle. Leptons and jets are merged with 2 layers of 40 nodes before they are merged with the $\met$ features. Finally, 4 layers of 180 nodes are added.
Fig.~\ref{fig:roc} shows the receiver operating characteristics (ROC) curve, which has been widely used as a measure of performance. From Fig.~\ref{fig:roc}, improved performance can be found for the particle-based model compared to the dense model. Studies using a DNN model including low and high-level features have been performed, but no significant improvement was found.

Similar studies have been performed using a BDT, with 1000 trees of 5 maximum depth, and `Adaptive Boost' algorithm. 
Fig.~\ref{fig:roc} shows the performance of each discriminant variable. Calculated area under curve (AUC) is 0.788, 0.762, 0.776, 0.666, and 0.591 for the particle-based DNN, dense DNN, BDT,  $p_T^{~l_1}$, and $\Delta\phi_{\text{jj}}$, respectively. The DNN particle-based model has slightly more discriminative power comparing to BDT, and is much more powerful than the single variables, $p_T^{~l_1}$ and $\Delta\phi_{\text{jj}}$.

\begin{figure}[htbp]
  \begin{center}
  \includegraphics[width=0.4\textwidth]{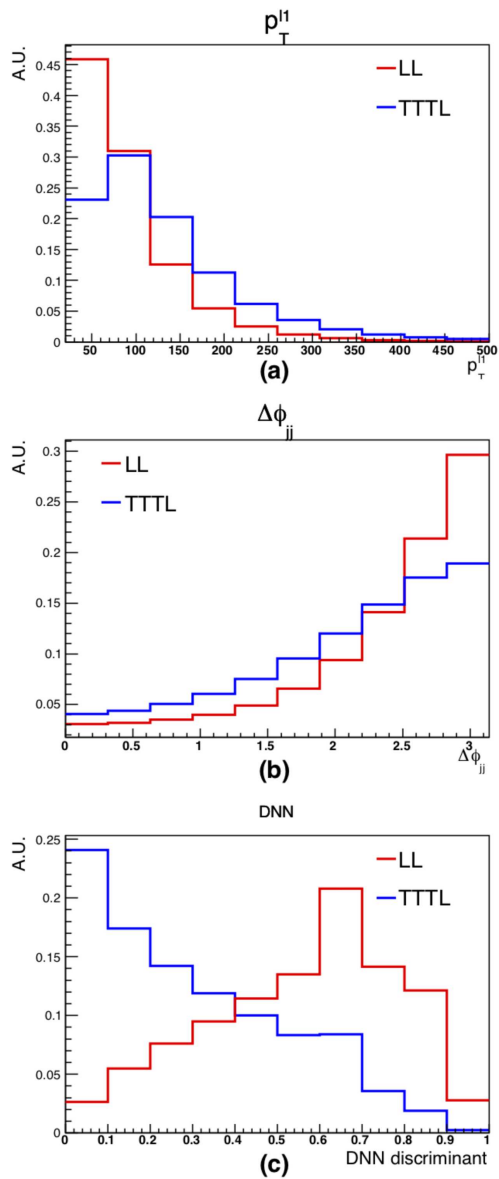}
    \caption{Kinematic distributions for $p_T^{~l_1}$ (a), $\Delta\phi_{\text{jj}}$ (b), and DNN discriminant corresponding to the particle-based model (c). }
    \label{fig:vdist}
  \end{center}
\end{figure}

Fig.~\ref{fig:vdist} shows several kinematic distributions of DNN inputs, and the distribution for the DNN discriminant itself. One can clearly see that the DNN greatly improves signal-background discrimination compared to rectangular cuts.

\begin{figure}[htbp]
  \begin{center}
   \includegraphics[width=0.4\textwidth]{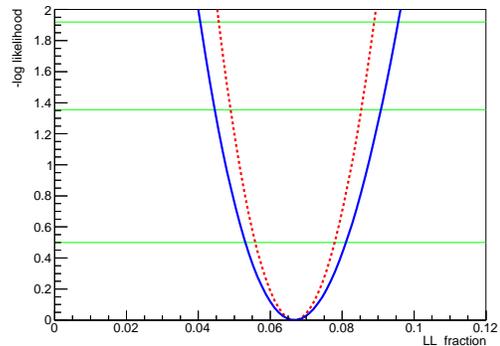}
    \caption{Fit results for the LL fraction using particle-based DNN in Mjj${>}$1500 GeV region. The horizontal lines represent the 68\%, 90\%, and 95\% confidence level, from lower to upper. Apart from those, the solid and dotted lines are the log likelihood distributions, with or without systematics included.}
    \label{fig:HistFactory}
  \end{center}
\end{figure}

We perform a fit to the DNN output and extract the LL fraction. The estimated LL fraction and accuracy is calculated by applying 2\% luminosity uncertainty, and 5\% systematic uncertainty both on LL and TT+TL. Table~\ref{tab:fraction} shows the DNN results compared with methods based on $p_T^{~l_1}$ and $\Delta\phi_{\text{jj}}$. Examples of fit results can be seen in Fig.~\ref{fig:HistFactory}, for $m_{\text{jj}}>1500$~GeV, which are achieved by HistFactory~\cite{histfactory} and cross-checked with RooFit~\cite{roofit}.

\begin{table}[htbp]
\begin{center}
{\footnotesize
\begin{tabular}{|c|c|c|c|c|c|}
  \hline 
 $m_{\text{jj}}$ cut & True Fraction & $p_T^{~l_1}$ & $\Delta\phi_{\text{jj}}$ & DNN  \\
 \hline
 $>$~850 GeV & 6.66\% & $6.67\%^{+1.95\%}_{-1.90\%}$ & $6.67\%^{+2.80\%}_{-2.76\%}$ & $6.66\%^{+1.11\%}_{-1.04\%}$  \\
 $>$~1200 GeV & 6.68\% & $6.70\%^{+2.26\%}_{-2.22\%}$ & $6.70\%^{+3.29\%}_{-3.25\%}$ & $6.68\%^{+1.26\%}_{-1.20\%}$  \\
 $>$~1500 GeV & 6.67\% & $6.71\%^{+2.62\%}_{-2.57\%}$ & $6.68\%^{+3.85\%}_{-3.80\%}$ & $6.67\%^{+1.44\%}_{-1.37\%}$  \\
$>$~1800 GeV & 6.69\% & $6.70\%^{+3.02\%}_{-2.96\%}$ & $6.68\%^{+4.48\%}_{-4.42\%}$ & $6.69\%^{+1.63\%}_{-1.56\%}$  \\ 
 $>$~2000 GeV & 6.66\% & $6.67\%^{+3.34\%}_{-3.27\%}$ & $6.66\%^{+4.98\%}_{-4.93\%}$ & $6.66\%^{+1.79\%}_{-1.71\%}$  \\
  \hline
\end{tabular}
}
\end{center}
\caption{Fit results for LL fraction with varaious $m_{\text{jj}}$ cuts, at 68\% confidence level. True fraction is the LL fraction computed at generator level.}
\label{tab:fraction}
\end{table}

\begin{figure}[htbp]
  \begin{center}
   \includegraphics[width=0.4\textwidth]{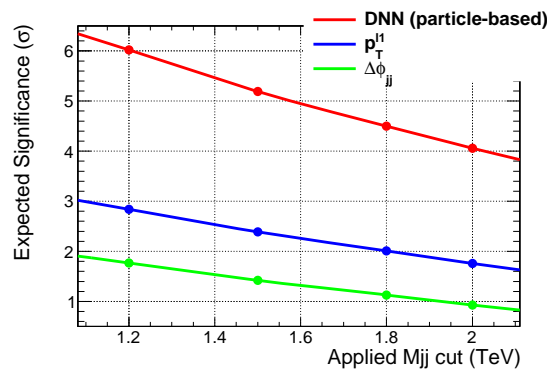}
    \caption{Significance dependence on $m_{\text{jj}}$ selection, through fitting $p_T^{~l_1}$, $\Delta\phi_{\text{jj}}$, or the DNN discriminant. Greatly improved performance is found with DNN.}
    \label{fig:significance}
  \end{center}
\end{figure}

Finally, we report here the significance. As mentioned above, the VBS $W^\pm W^\pm$ process profits from lower background than in other VBS channels, considering that dominant backgrounds (WZ and hadron decays) are greatly suppressed and asymptotically negligible at high $m_{\text{jj}}$~\cite{Sirunyan:2017ret,ATLAS:2018ogo,CMS-PAS-FTR-18-005}. On the other hand, contributions from those dominant backgrounds can be estimated in experimental analysis and thus subtracted keeping uncertainties under control.
In the ranges $m_{\text{jj}}>$ 1500 and 2000 GeV, significances of 5.2 and 4.1 standard deviations can be achieved from a likelihood fit of DNN distributions. The same study has been performed via $p_T^{~l_1}$ and $\Delta\phi_{\text{jj}}$. Fig.~\ref{fig:significance} shows greatly improved significance obtained with DNN.

In summary, measuring the longitudinally polarized fraction of $W^\pm W^\pm$ scattering at the LHC is crucial to examine the unitarization mechanism of the vector boson scattering amplitude through Higgs and possible new physics. We apply here for the first time a Deep Neural Network classification to extract the longitudinal fraction. Based on fast simulation implemented with the Delphes framework, significant improvement from DNN is found to be achievable and robust.  An observation with an integrated luminosity of 3000 $fb^{-1}$ is found to reach 4 standard deviations at high $m_{\text{jj}}$ region, such as above 2 TeV, where backgrounds are negligible. With a combination of the CMS and ATLAS measurements at the HL-LHC, an observation above 5 standard deviations can be expected with the Deep Learning technique proposed in this study.

\pagebreak
\newpage
\acknowledgements
This work is supported in part by the National Natural Science Foundation of China, under Grants No. 11175251 and 11205008, by MOST under grant No. 2018YFA0403900, and COST Action CA16108.
We thank the CNRS/IN2P3 and the France China Particle Physics Laboratory (FCPPL) for their support. We also thank Ga\"{e}l Touquet for his interesting suggestions about DNN, and Junjie Zhu and Pietro Govoni for their helpful discussions.

\bibliographystyle{ieeetr}
\bibliography{h}

\begin{thebibliography}{10}

\bibitem{Sirunyan:2017ret} 
  A.~M.~Sirunyan {\it et al.} [CMS Collaboration],
  Phys.\ Rev.\ Lett.\  {\bf 120}, no. 8, 081801 (2018)
  doi:10.1103/PhysRevLett.120.081801
  [arXiv:1709.05822 [hep-ex]].

\bibitem{ATLAS:2018ogo} 
  The ATLAS collaboration [ATLAS Collaboration],
  ATLAS-CONF-2018-030.

\bibitem{WWuni}
M.~J.~G.~Veltman, Acta Phys. Polon. B8 (1977) 475; B.~W.~Lee, C.~Quigg, and H.~B.~Thacker, Phys. Rev. Lett. 38 (1977) 883-885; B.~W.~Lee, C.~Quigg, and H.~B.~Thacker, Phys. Rev. D16 (1977) 1519.

\bibitem{Alboteanu:2008my} 
  A.~Alboteanu, W.~Kilian and J.~Reuter,
  JHEP {\bf 0811}, 010 (2008)
  doi:10.1088/1126-6708/2008/11/010
  [arXiv:0806.4145 [hep-ph]].

\bibitem{Chang:2013aya} 
  J.~Chang, K.~Cheung, C.~T.~Lu and T.~C.~Yuan,
  Phys.\ Rev.\ D {\bf 87}, 093005 (2013)
  doi:10.1103/PhysRevD.87.093005
  [arXiv:1303.6335 [hep-ph]].

\bibitem{Doroba12} 
K.~Doroba, J.~Kalinowski, J.~Kuczmarski, S.~Pokorski, J.~Rosiek, et al., Phys.\ Rev.\ D {\bf 86} (2012) 036011. 
[arXiv:1201.2768 [hep-ph]].

\bibitem{Freitas:2012uk} 
  A.~Freitas and J.~S.~Gainer,
  Phys.\ Rev.\ D {\bf 88}, no. 1, 017302 (2013)
  doi:10.1103/PhysRevD.88.017302
  [arXiv:1212.3598 [hep-ph]].

\bibitem{Searcy:2015apa} 
  J.~Searcy, L.~Huang, M.~A.~Pleier and J.~Zhu,
  Phys.\ Rev.\ D {\bf 93}, no. 9, 094033 (2016)
  doi:10.1103/PhysRevD.93.094033
  [arXiv:1510.01691 [hep-ph]].

\bibitem{CMS-PAS-SMP-14-008}
{CMS} Collaboration, {\em {Prospects for the study of vector boson scattering
  in same sign WW and WZ interactions at the HL-LHC with the upgraded CMS
  detector}\/},  CMS Physics Analysis Summary (2016) no.~CMS-PAS-SMP-14-008, .
  \url{https://cds.cern.ch/record/2220831}.

\bibitem{CMS-PAS-FTR-18-005}
{CMS} Collaboration, {\em {Study of $W^\pm W^\pm$ production via vector boson scattering at the HL-LHC with the upgraded CMS detector}\/},  CMS Physics Analysis Summary (2018) no.~CMS-PAS-FTR-18-005, .
  \url{https://cds.cern.ch/record/2646870}.

\bibitem{Keras} 
F.~Chollet et al., https://github.com/fchollet/keras

\bibitem{Tensorflow} 
Martín Abadi et al., TensorFlow: Large-scale machine learning on heterogeneous systems,
2015. Software available from tensorflow.org.

\bibitem{Baldi:2014kfa} 
  P.~Baldi, P.~Sadowski and D.~Whiteson,
  Nature Commun.\  {\bf 5}, 4308 (2014)
  doi:10.1038/ncomms5308
  [arXiv:1402.4735 [hep-ph]].

\bibitem{Baldi:2014pta} 
  P.~Baldi, P.~Sadowski and D.~Whiteson,
  Phys.\ Rev.\ Lett.\  {\bf 114}, no. 11, 111801 (2015)
  doi:10.1103/PhysRevLett.114.111801
  [arXiv:1410.3469 [hep-ph]].

\bibitem{Roe:2004na} 
  B.~P.~Roe, H.~J.~Yang, J.~Zhu, Y.~Liu, I.~Stancu and G.~McGregor,
  Nucl.\ Instrum.\ Meth.\ A {\bf 543}, no. 2-3, 577 (2005)
  doi:10.1016/j.nima.2004.12.018
  [physics/0408124].

\bibitem{TMVA} 
 Hocker, A. et al. TMVA - Toolkit for Multivariate Data Analysis. PoS ACAT, 040 (2007).

\bibitem{Alwall:2014hca} 
  J.~Alwall {\it et al.},
  JHEP {\bf 1407}, 079 (2014)
  doi:10.1007/JHEP07(2014)079
  [arXiv:1405.0301 [hep-ph]].

\bibitem{Sjostrand:2003wg} 
  T.~Sjostrand, L.~Lonnblad, S.~Mrenna and P.~Z.~Skands,
  hep-ph/0308153.

\bibitem{deFavereau:2013fsa} 
  J.~de Favereau {\it et al.}  [DELPHES 3 Collaboration],
  JHEP {\bf 1402}, 057 (2014)
  [arXiv:1307.6346 [hep-ex]].

\bibitem{Rainwater:1996ud} 
  D.~L.~Rainwater, R.~Szalapski and D.~Zeppenfeld,
  Phys.\ Rev.\ D {\bf 54}, 6680 (1996)
  doi:10.1103/PhysRevD.54.6680
  [hep-ph/9605444].

\bibitem{histfactory} 
K.~Cranmer, G.~Lewis, L.~Moneta, A.~Shibata, W.~Verkerke, Hist-Factory: a tool for creating statistical models for use with RooFit
and RooStats. CERN-OPEN-2012-016 (2012).  https://cds.cern.ch/record/1456844/

\bibitem{roofit}
W.~Verkerke, D.~P.~Kirkby, The RooFit toolkit for data modeling, in: L.~Lyons, M.~Karagoz (Eds.), Proceedings of PHYSTAT 05: Statistical Problems in Particle Physics,
Astrophysics and Cosmology, Oxford, England, 2005, arXiv:physics/0306116. 


\end{thebibliography}
\end{document}